\documentclass[preprint]{elsarticle}
\usepackage{bm}
\usepackage[T2A]{fontenc}
\usepackage[cp1251]{inputenc}
\usepackage[english,russian]{babel}
\usepackage{amsfonts,amssymb,amsmath}
\usepackage{amsmath,graphicx}
\usepackage{graphicx,pifont} 
\usepackage{epsfig}
\title{Exact equation for classical many-particle systems in closed form: from mechanics to statistical thermodynamics}
\author[ayz]{A.Yu.~Zakharov} 
\ead{Anatoly.Zakharov@novsu.ru}
\address{Yaroslav-the-Wise Novgorod State University, Velikiy Novgorod, 173003, Russia}
 \begin{document}
\begin{abstract}
The exact equations of motion for microscopic density of classical particles number with account of inter-particle interactions and external field in closed form are derived. An integral equation for equilibrium distributions of the particles is deduced. No statistical or probabilistic hypotheses and assumption in these deductions have been used. Some well known results of equilibrium statistical mechanics are deduced from the obtained equations as simple limiting cases. The wave equation for almost homogeneous systems with inter-particle interactions are obtained. Connection between inter-particle potential and dispersion law of sound is established.
\end{abstract}


\maketitle

\section{introduction}
At present, there are two basic approaches to classical statistical mechanics.
\begin{enumerate}
\item The Gibbs' ensembles approach. In this method there is the difficult problem of the partitions functions evaluation. There is no methods of the problem partition functions for general form inter-atomic potential calculation. Unfortunately, inaccuracy of all existing methods of partition functions approximate evaluation have not a priory estimated. 
\item The Bogoliubov-Born-Green-Kirkwood-Yvon (BBGKY) distribution functions method. In this method there is most essential difficulty~--- the closing problem of nonclosed system equations. All the existing methods of this equations closure contain some additional inpredicable inaccuracies. 
\end{enumerate}
In addition, there are numerous efforts of descriptions many-body systems by means various versions of kinetic equations from Boltzmann equation (including the Vlasov equation with self-consistent field) to quantum kinetic equations (such as Kadanoff-Baym equations and Keldysh equations). But, these kinetic equations application to concrete problems have some similar difficulties as pure equilibrium approaches.

The aim of the present paper is derivation of exact closed equation for many particle classical system evolution. This derivation based on the Newtonian laws only and does not based on any probabilistic or statistical assumptions and hypotheses. Some of consequences this equation are derived.

\section{First order equation}

Let us define the microscopic local density~$n(\mathbf{r},t)$ of a classical system by relation
\begin{equation}\label{n(r,t)}
n(\mathbf{r},t) = \sum_{s=1}^N\ \delta\left(\mathbf{r} - \mathbf{R}_s(t) \right) = \int\, \frac{d\mathbf{k}}{\left( 2\pi\right)^3 }\,e^{i\,\mathbf{k}\,\mathbf{r}} \ \tilde{n}(\mathbf{k},t),
\end{equation}
where~$\tilde{n}(\mathbf{k},t)$ are the collective coordinates
\begin{equation}\label{tilden(k,t)}
\tilde{n}(\mathbf{k},t) = \sum_s e^{-i\,\mathbf{k}\,\mathbf{R}_s(t)}.
\end{equation}
These coordinates were first used by Bohm and Pines in paper~\cite{BP}.

Evaluation of the sums of type~$ \sum_s \ f \left (\mathbf {R} _s \right) $, where~$ f \left (\mathbf {R} _s \right) $ are any ``the one-particle'' functions, will carried out by a rule
\begin{equation}\label{sum-over-s}
\sum_s \ f\left( \mathbf{R}_s(t)\right) = \sum_s \int\ f\left( \mathbf{r}\right)\, \delta \left(\mathbf{r} - \mathbf{R}_s(t) \right)\, d\mathbf{r} = \int\, f\left( \mathbf{r}\right)\, n\left(\mathbf{r}, t \right)\, d\mathbf{r}. 
\end{equation}  

The first derivative of a microscopic density with respect to time is related to the instantaneous velocities of particles. All the interactions both between particles and particles with external field can be manifested in the second derivative with respect to time~$t$.  Therefore it should be expected, that for the evolution of system description (at least without retarding of interactions) it is enough to use the equations containing derivatives with respect to time not above the second order.

After differentiating the local density~$n(\mathbf{r},t)$ with respect to time, we have 
\begin{equation}\label{dn/dt}
\dfrac{\partial n(\mathbf{r},t)}{\partial t} = -i \int\, \frac{d\mathbf{k}}{\left( 2\pi\right)^3 }\,e^{i\,\mathbf{k}\,\mathbf{r}} \ \sum_s e^{-i\,\mathbf{k}\,\mathbf{R}_s(t)} \left( \mathbf{k}\cdot \dfrac{d\mathbf{R}_s(t) }{dt} \right) 
\end{equation}
Using~(\ref{sum-over-s}) and Fourier representation of delta-function, we obtain  
\begin{equation}
\dfrac{\partial n(\mathbf{r},t)}{\partial t} = \ \int\ \left( \dfrac{\partial \delta\left(\mathbf{r} - \mathbf{R} \right)}{\partial \mathbf{R}} \cdot \mathbf{v}\left(\mathbf{R}, t\right)\right)  \, n\left(\mathbf{R},t \right)\, d\mathbf{R} = \ -\nabla\left(\mathbf{v}\left(\mathbf{r}, t \right) \, n\left(\mathbf{r}, t \right) \right).
\end{equation}
This is none other than well known the equation of continuity, i.e. law of the particles number in system conservation.
\begin{equation}\label{contin}
\dfrac{\partial n(\mathbf{r},t)}{\partial t} + \nabla\left(\mathbf{v}\left(\mathbf{r},t \right) \, n\left(\mathbf{r}, t \right) \right) =0.
\end{equation}
Let's pass to derivation of a second order equation.

\section{Second order equation}

Let us a second derivative of function~$ n (\mathbf{r}, t) $ with respect to time
\begin{equation}\label{ddot-n}
\dfrac{\partial^2 n(\mathbf{r},t)}{\partial t^2} = \int\, \frac{d\mathbf{k}}{\left( 2\pi\right)^3 }\,e^{i\,\mathbf{k}\,\mathbf{r}} \ \sum_s e^{-i\,\mathbf{k}\,\mathbf{R}_s(t)}\left\lbrace -  \left(\mathbf{k} \cdot \dot{\mathbf{R}}_s (t) \right)^2 -i\left( \mathbf{k}\cdot \ddot{\mathbf{R}}_s(t)\right)  \right\rbrace. 
\end{equation}
The first summand in the integrand of this expression can be transformed to the following integral
\begin{equation}\label{dot-R}
-\sum_s e^{-i\,\mathbf{k}\,\mathbf{R}_s(t)}\, \left(\mathbf{k} \cdot \dot{\mathbf{R}}_s (t) \right)^2\ = \ -\frac{k^2}{D}\, \int\, e^{-i \mathbf{kR}}\, n(\mathbf{R},t)\,v^2(\mathbf{R},t)\, d\mathbf{R},
\end{equation}
where~$D$ is the space dimensionality.

To calculate the second summand in the integrand~(\ref{ddot-n}), we should evaluate~$\ddot{\mathbf{R}}_s$. According to the second Newton's law, we have:
\begin{equation}\label{ddot-Rs}
\begin{array}{l}
{\displaystyle \ddot{\mathbf{R}}_s(t)\ =\ -\frac 1m\ \nabla_{\mathbf{R}_s} \left[  \sum_{s'} W\left(\mathbf{R}_s(t) - \mathbf{R}_{s'}(t) \right)  + \varphi\left(\mathbf{R}_s,\, t\right) \right] }\\
{\displaystyle  =\ -\frac 1 m \nabla_{\mathbf{R}_s} \left[  \int W\left(\mathbf{R}_s(t) - \mathbf{R}' \right) n\left(\mathbf{R}',t \right)\, d\mathbf{R}' +  \varphi\left(\mathbf{R}_s,\, t\right)   \right]  }\\
\end{array}
\end{equation}
where~$W\left(\mathbf{R}_s - \mathbf{R}_{s'}\right)$ is an interaction potential between particles located in points~$\mathbf{R}_s$ and $\mathbf{R}_{s'}$, $\varphi\left(\mathbf{r},\, t \right) $~is an external field potential, $m$~is a particle mass. 

Substituting~(\ref{ddot-Rs}) into the second summand in integrand~(\ref{ddot-n}), we obtain 
\begin{equation}\label{ddot-R}
\begin{array}{l}
{\displaystyle i\sum_s e^{-i\,\mathbf{k}\,\mathbf{R}_s(t)}\, \left( \mathbf{k}\cdot \ddot{\mathbf{R}}_s(t) \right) }\\
{\displaystyle =\ -\frac{i}{m}\, \int e^{-i \mathbf{k} \mathbf{R}}\, n(\mathbf{R},t) \biggl[  \left(\mathbf{k}\cdot \nabla_{\mathbf{R}} \int W\left(\mathbf{R} - \mathbf{R}' \right) n\left(\mathbf{R}',t \right)\, d\mathbf{R}'  \right) }\\
{\displaystyle 
 +     \left(\mathbf{k}\cdot \nabla_{\mathbf{R}}\,  \varphi\left(\mathbf{R}, t \right) \right)     \biggr] d\mathbf{R}  }
\end{array}
\end{equation}
Using~(\ref{dot-R}), (\ref{ddot-Rs}), and~(\ref{ddot-n}) leads to following results
\begin{equation}\label{ddot-n1}
\begin{array}{l}
{\displaystyle -\int\, \frac{d\mathbf{k}}{\left( 2\pi\right)^3 }\,e^{i\,\mathbf{k}\,\mathbf{r}} \ \sum_s e^{-i\,\mathbf{k}\,\mathbf{R}_s(t)}\left\lbrace \left(\mathbf{k} \cdot \dot{\mathbf{R}}_s (t) \right)^2 \right\rbrace =\, \frac{1}{D}\int \dfrac{\partial^2 \delta \left(\mathbf{r} - \mathbf{R} \right) }{\partial \mathbf{r}^2}\, n(\mathbf{R},t)\,v^2(\mathbf{R},t)\, d\mathbf{R} }\\
{\displaystyle =\, \frac{1}{D}\Delta\left[ n(\mathbf{r},t)\,v^2(\mathbf{r},t)\right]  }
\end{array}
\end{equation}
($\Delta$~ is the Laplace operator)
and
\begin{equation}\label{ddot-n2}
\begin{array}{l}
{\displaystyle \int\, \frac{d\mathbf{k}}{\left( 2\pi\right)^3 }\,e^{i\,\mathbf{k}\,\mathbf{r}} \ \sum_s e^{-i\,\mathbf{k}\,\mathbf{R}_s(t)}\left\lbrace i\left( \mathbf{k}\cdot \ddot{\mathbf{R}}_s(t)\right)  \right\rbrace }\\
{\displaystyle =\,  - \frac{1}{m} \nabla_{\mathbf{r}}  \left[ n(\mathbf{r},t) \left( \nabla_{\mathbf{r}} \left\lbrace  \int W\left(\mathbf{r} - \mathbf{R} \right) n\left(\mathbf{R},t \right)\, d\mathbf{R} + \varphi\left(\mathbf{r}, t \right) \right\rbrace  \right) \right] }\\
\end{array}
\end{equation}

We shall return to~(\ref{ddot-n1}), namely to local value of~$v^2(\mathbf{r})$ at the point~$\mathbf{r}$. This value in classical kinetics related to (local) absolute temperature
\begin{equation}\label{mean-v2}
\frac{m\,v^2(\mathbf{r},t)}{2}\ =\ \frac{D}{2}\,\kappa T(\mathbf{r},t), 
\end{equation} 
where~$\kappa$~is the Boltzmann constant, $T(\mathbf{r})$~is the absolute temperature.  We shall assume in this paper that~$ v^2 (\mathbf {r}) $ does not depend on the coordinates~$\mathbf {r}$ and time~$t$.

Substituting expressions~(\ref{ddot-n1}), (\ref{ddot-n2}) into~(\ref{ddot-n}), we obtain the basic equation 
\begin{equation}\label{equat-bas}
\begin{array}{r}
{\displaystyle \dfrac{\partial^2n(\mathbf{r},t) }{\partial t^2}\, = \, \frac{v^2}{D}\,\Delta\left[ n(\mathbf{r},t)\right]  }\\ \\
{\displaystyle + \frac{1}{m} \nabla_{\mathbf{r}}  \left[ n(\mathbf{r},t) \left( \nabla_{\mathbf{r}} \left\lbrace  \int W\left(\mathbf{r} - \mathbf{R} \right) n\left(\mathbf{R},t \right)\, d\mathbf{R} + \varphi\left(\mathbf{r},\, t \right) \right\rbrace  \right) \right]}.
\end{array}
\end{equation}
The further part of the present paper is devoted to the analysis of this equation and its consequences.

\section{Statical solutions of the basic equation}

Let us consider at first the problem of equilibrium solutions for the basic equation at not depending on time external field, i.e. 
\begin{equation}\label{equilibr}
\varphi =   \varphi(\mathbf{r}); \quad \dfrac{\partial^2n(\mathbf{r},t) }{\partial t^2}\, \equiv 0.
\end{equation}
Then we have
\begin{equation}\label{n(r)}
\frac{v^2}{D}\,\Delta\left[ n(\mathbf{r})\right]\, + \frac{1}{m} \nabla_{\mathbf{r}}  \left[ n(\mathbf{r}) \left( \nabla_{\mathbf{r}} \left\lbrace  \int W\left(\mathbf{r} - \mathbf{R} \right) n\left(\mathbf{R} \right)\, d\mathbf{R} + \varphi\left(\mathbf{r}\right) \right\rbrace  \right) \right] =\, 0.
\end{equation}
After partial integration we obtain  
\begin{equation}\label{n(r)-stat}
\frac{v^2}{D}\, \nabla n(\mathbf{r})  + \, \frac{1}{m}   n(\mathbf{r})\  \nabla_{\mathbf{r}} \left\lbrace  \int W\left(\mathbf{r} - \mathbf{R} \right) n\left(\mathbf{R} \right)\, d\mathbf{R} + \varphi\left(\mathbf{r}\right) \right\rbrace \, =\, \mathbf{A},
\end{equation}
where~$\mathbf{A} $ is a constant vector.

In particular, at~$\mathbf{A}=0$ with account condition~(\ref{mean-v2}) it leads to the integral equation
\begin{equation}\label{equil}
n(\mathbf{r}) = C\ \exp{\left[ -\,\frac{1}{\kappa T}\left\lbrace  \int W\left(\mathbf{r} - \mathbf{R} \right) n\left(\mathbf{R} \right)\, d\mathbf{R} + \varphi\left(\mathbf{r}\right) \right\rbrace \right] }
\end{equation}
($C$ is a constant). This equation have a form of the Boltzmann distribution with some {\em effective field} consisting of an external field~$\varphi\left(\mathbf{r}\right) $  and a local field 
\begin{equation}\label{local-field}
\tilde{\varphi}(\mathbf{r}) = \int W\left(\mathbf{r} - \mathbf{R} \right) n\left(\mathbf{R} \right)\, d\mathbf{R},
\end{equation}
due to interactions between the particles.  

In absence of an external field~(i.e. at~$\varphi\left(\mathbf{r}\right) \equiv 0$) this equation has the following form
\begin{equation}\label{Vlasov}
n(\mathbf{r}) = C\ \exp{\left[ -\,\frac{1}{\kappa T}\left\lbrace  \int W\left(\mathbf{r} - \mathbf{R} \right) n\left(\mathbf{R} \right)\, d\mathbf{R} \right\rbrace \right] }.
\end{equation}
This equation was at first derived by Vlasov~\cite{Vlasov} from the collisionless Boltzmann equation by using some additional hypotheses. Later the same equation was derived by Bazarov~\cite{Baz1} from BBGKY hierarchy with assuming the multiplicative closure of this hierarchy.

\section{The analysis of basic equation solution in an almost homogeneous system}
Let us consider the basic equation~(\ref{equat-bas}) for almost homogeneous system without external field. In this case a solution can be presented in the following form
\begin{equation}\label{n1}
n(\mathbf{r},t) = n_0 + n_1(\mathbf{r},t), \quad (n_0 = \mathrm{const},\  \left| n_1(\mathbf{r},t)\right| \ll n_0). 
\end{equation}
Linearization of the basic equation over~$n_1(\mathbf{r},t)$ at indicated conditions leads to following linear equation
\begin{equation}\label{lin}
\dfrac{\partial^2n_1(\mathbf{r},t) }{\partial t^2}\, = \, \frac{1}{m} \left[  \kappa T\, \Delta n_1(\mathbf{r},t) + {n_0}\, \Delta \int\, W\left(\mathbf{r} - \mathbf{R} \right)\, n_1(\mathbf{R},t)\, d\mathbf{R} \right] .  
\end{equation}

As we can see, the interactions between particles lead to substantial distortion of the wave equation form: instead of an usual hyperbolic partial equation we have an integro-differential equation with the integral term of convolutional type.

The Cauchy problem for linearized equation~(\ref{lin}) has the following form
\begin{equation}\label{Cauchy}
\left\lbrace 
\begin{array}{l}
{\displaystyle n_1(\mathbf{r}, 0) = f_1(\mathbf{r});  }\\
{\displaystyle  \dot{n}_1(\mathbf{r},0) = f_2(\mathbf{r})  }
\end{array}
\right..
\end{equation}
Its solution is quite elementary. Let us introduce the Fourier-transform of the function~$n_1(\mathbf{r},t)$
\begin{equation}\label{n1(k,t)}
n_1(\mathbf{r},t)\ =\ \int\, \frac{d\mathbf{k}}{\left( 2\pi\right)^3 }\ e^{i\,\mathbf{k}\,\mathbf{r}} \, \tilde{n}_1(\mathbf{k},t)
\end{equation}
and obtain the following equation for function~$\tilde{n}_1(\mathbf{k},t)$
\begin{equation}\label{n1(k,t)-equat}
\dfrac{\partial^2\tilde{n}_1(\mathbf{k},t) }{\partial t^2}\, + \ \frac{k^2}{m}\, \left[\kappa T\, +\, {n_0}\ \widetilde{W}(\mathbf{k}) \right]\tilde{n}_1(\mathbf{k},t)\,  =\, 0.
\end{equation}
This equation has as a consequence the following dispersion law of oscillations
\begin{equation}\label{dispersion}
\omega^2\, =\, \frac{k^2}{m}\, \left[\kappa T\, +\, {n_0}\ \widetilde{W}(\mathbf{k}) \right], 
\end{equation}
where~$ \widetilde{W}(\mathbf{k})$~is a Fourier-transform of inter-atomic potential~${W}(\mathbf{r})$ (note that for a central potential both of the functions~${W} (\mathbf{r})$ and  $\widetilde {W} (\mathbf{k}) $ depend on the magnitudes of their arguments). Let's suppose, that~$ \widetilde{W}(k)$ satisfies to Dobrushin-Ruelle-Fischer condition of thermodynamic stability of inter-atomic potentials
\begin{equation}\label{Ruelle}
\widetilde{W}(k) \geq 0.
\end{equation}

Hence we obtain the Cauchy problem~(\ref{lin},\ref{Cauchy}) solution:
\begin{equation}\label{sound}
n_1(\mathbf{r},t)\ =\ \int\, \frac{d\mathbf{k}}{\left( 2\pi\right)^3 }\ e^{i\,\mathbf{k}\,\mathbf{r}} \, \left[ C_1(\mathbf{k}) e^{i\, \omega(k)\,t  } + C_2(\mathbf{k}) e^{-i\, \omega(k)\,t  } \right], 
\end{equation}
where~$C_1(\mathbf{k})$ and~$C_2(\mathbf{k})$~are the functions, determined by conditions~(\ref {Cauchy}):
\begin{equation}\label{Ci(k)}
\left\lbrace 
\begin{array}{l}
{\displaystyle C_1(\mathbf{k}) \, = \,\frac{1}{2}\, \left[ F_1(\mathbf{k})\, - \, \frac{i}{\omega(\mathbf{k})} F_2(\mathbf{k})  \right] ;  }\\
{\displaystyle  C_2(\mathbf{k}) \, = \, \frac{1}{2}\,\left[ F_1(\mathbf{k})\, + \, \frac{i}{\omega(\mathbf{k})} F_2(\mathbf{k})  \right],   }
\end{array}
\right.
\end{equation}
$F_1(\mathbf{k})$ and~$F_2(\mathbf{k})$ are the Fourier-transforms functions $f_1(\mathbf{k})$ and $f_2(\mathbf{r})$, respectively.

It should be noted especially, that the equation~(\ref{dispersion}) contains the a possibility to determine the real inter-atomic potential~${W}(\mathbf{r})$ under the dispersion law~$\omega(\mathbf{k})$.

\section{Conclusion}

The papaer contains the following results.

\begin{itemize}
\item Derivation of exact equation of motion for microscopic density of classical many-body system consisting of interacting particles. This derivation does not based on any statistical or probabilistic hypotheses and assumption.
\item Some well known basic foundation of equilibrium statistical mechanics are deduced from the obtained equations as simple limiting cases.
\item The wave equation for many-body system with inter-particle interactions is obtained. Connection between inter-particle potential and dispersion law of sound is established.
\end{itemize}

In this context there are some open problems. One of the most essential of these problems is problem of irreversibility within the proposed approach. There is no the full clearness in connection between the offered equation~(\ref{equat-bas}) and classical non-equilibrium statistical mechanics. In contrast to equations of motion in non-equilibrium classical statistical mechanics, exact equation~(\ref{equat-bas}) contain derivatives of second order with respect to time. Therefore the solutions of equation~(\ref{equat-bas}) does not contain passage to the equilibrium state of a system at $t \to \infty $. Thus, there is a problem: how to adapt the {\em exact} equation(\ref{equat-bas}) for irreversible processes description. There are number ways for this problem solution. By any way it should be based on probabilistic arguments.

\section{Acknowledgements}
This work is fulfilled by partial financial support Russian Ministry of Education and Science within the framework of a base part.

\end{document}